\documentclass[12pt]{revtex4-1}
\usepackage{amsfonts,amsmath,amssymb,xcolor}
\usepackage{amsthm,amscd,ulem,bm}
\usepackage{graphicx}
\makeatletter

\@addtoreset{equation}{section}
\makeatother
\theoremstyle{definition}

\newcommand{\bib}[2]{\frac{\partial {#1}}{\partial {#2}}}
\newcommand{\bbib}[3]{\frac{\partial^2 {#1}}{\partial {#2}{\partial {#3}}}}

\def\b1{{ {\bf 1}}}

\def\therefore{%
\setlength{\unitlength}{1pt}%
\thinlines %
\begin{picture}(10,10)%
\put(0, 0){.}
\put(3, 6){.}
\put(6,0){.}
\end{picture}%
}%
\begin{document}
\title{
Killing Symmetry on Finsler Manifold
}
\author{Takayoshi Ootsuka and Ryoko Yahagi}
\email{ootsuka@cosmos.phys.ocha.ac.jp}
\email{yahagi@hep.phys.ocha.ac.jp}
\affiliation{Physics Department, Ochanomizu University, 2-1-1 Ootsuka Bunkyo, Tokyo, Japan }
\author{Muneyuki Ishida}
\email{ishida@phys.meisei-u.ac.jp}
\affiliation{Department of Physics, Meisei University, 2-1-1 Hodokubo Hino, Tokyo191-8506, Japan}
\date{\today}

\begin{abstract} 
Killing vector fields $K$ are defined on Finsler manifold.
The Killing symmetry is reformulated  simply as  
 $\delta K^\flat =0$ by using  
the Killing non-linear 1-form $K^\flat$ and the spray operator $\delta$ 
with the Finsler non-linear connection.
$K^\flat$ is related to the generalization of Killing tensors on Finsler manifold, and 
the condition $\delta K^\flat =0$ 
gives an analytical method of finding higher derivative conserved quantities, 
which may be called hidden conserved quantities. 
We show two examples:
the Carter constant on Kerr spacetime and the Runge-Lentz vectors
in Newtonian gravity.

\end{abstract}


\maketitle

\section{Introduction}

Recently there arouse a number of interests on Finsler geometry, and  
it has a great variety of physical applications, such as special and general relativity~\cite{Gibbons}, classical mechanics~\cite{Lanc}, path integral~\cite{OT}, field theories~\cite{OF}, 
string duality~\cite{YS}, thermodynamics, fluid dynamics~\cite{fluid} and so on.  
A Finsler manifold $(M,F)$ is a natural extension of a Riemannian manifold.
$M$ is a $(n+1)$-dimensional differentiable manifold and $F=F(x,dx)$ is a Finsler metric, 
which is a function of $(x^\mu, dx^\mu)$. 
Here the coordinates $x^\mu$ and 1-forms $dx^\mu$ are treated as coordinates of tangent bundle. 
We consider the subbundle $D(F) \subset TM$ where $F(x,dx)$ is well defined. 
$F(x,dx)$ satisfies the homogeneity relation of degree 1,
\begin{eqnarray}
F(x,\lambda dx)=\lambda\ F(x,dx),\ \ \lambda>0\ \  &\Leftrightarrow & dx^\mu \frac{\partial F}{\partial dx^\mu} = F\ .
\label{eq1}
\end{eqnarray}
A Finsler metric $F$ gives a Lagrangian of point particle and
its action is defined by the geometrical length of the oriented curve ${ c}$ on $M$ as 
\begin{eqnarray}
{\cal A}[{c}] &=& \int_{c} F(x,dx)\ .
\label{eq2}
\end{eqnarray}
The variational principle leads the Euler-Lagrange equation
\begin{eqnarray}
0 &=&
c^*\left\{
   \frac{\partial F}{\partial x^\mu} -d \left(  \frac{\partial F}{\partial dx^\mu} \right) 
 \right\}
\label{eq3}
\end{eqnarray}
on the solution curves ${c}$.
The above equation is rewritten in the geodesic form
\begin{eqnarray}
d^2 x^\mu + 2G^\mu (x,dx) = \lambda\ dx^\mu\ ,
\label{eq4}
\end{eqnarray}
where the Lagrange multiplier  $\lambda=\lambda(x,dx)$ comes from the homogeneity of $F$, Eq.~(\ref{eq1}).
The non-linear connection $N^\mu{}_\alpha$ (and $\nabla$) are defined~\cite{KO} by
\begin{eqnarray}
\nabla dx^\mu &=&  -dx^\alpha \otimes N^\mu{}_{\alpha}(x,dx) 
,\nonumber\\
\frac{\partial N^\mu{}_\alpha}{\partial dx^\beta} &-& \frac{\partial N^\mu{}_\beta}{\partial dx^\alpha}  =0,\ \ \  \ \ \frac{\partial F}{\partial x^\alpha} - N^\mu{}_\alpha\frac{\partial F}{\partial dx^\mu} =0\ 
\label{eq4-1}
\end{eqnarray}
which leads
\begin{eqnarray}
N^\mu{}_\alpha = \frac{\partial G^\mu}{\partial dx^\alpha}, &\ \ \ \ & 2 G^\mu = \left( dx^\beta \frac{\partial F}{\partial dx^\beta} \right) \frac{dx^\mu}{F} 
+ F^{ab} \l^\mu_a \left( -\frac{\partial F}{\partial x^b}+dx^\rho\frac{\partial^2 F}{\partial dx^b \partial x^\rho}  \right) \ ,\nonumber\\
&& a,b=1,\cdots,n;\ \ \ \ \ \ \ \ \mu,\alpha,\beta,\rho =0,1,\cdots,n\ ,
\label{eq4-2}
\end{eqnarray} 
where $F^{ab}$ is the inverse of the $n\times n$ space components of 
$F_{\mu\nu}=\frac{\partial^2 F}{\partial dx^\mu \partial dx^\nu} $ and $\l^\mu_a=\delta^\mu_a -\frac{p_adx^\mu}{F}$ with $p_a = \frac{\partial F}{\partial dx^a}$.
We note that in Riemannian manifold, taking $F=\sqrt{g_{\mu\nu}(x)dx^\mu dx^\nu}$, 
 $N^\mu{}_\alpha$ is given by the ordinary Christoffel symbol,
$N^\mu{}_\alpha =\varGamma^\mu{}_{\beta\alpha} dx^\beta$.  

In our previous work~\cite{OYIT} we have discussed the energy-momentum conservation
law on Finsler$/$Kawaguchi manifold. It is derived from the time variation
in Finsler$/$Kawaguchi space in conformity with the standard procedure of leading energy-momentum tensor. 
The conservation law is derived from a general background, the Killing symmetry or 
Noether's theorem, which is the aim of the present paper. 
Killing vector on Finsler manifold is defined in ref.\cite{Ootsuka}.
Here we formulate it covariantly in a simple form.
Furthermore, the Killing symmetry is given in another form. 
We define the Killing non-linear 1 form $K^\flat$ which directly defines the ``Killing tensors" 
on Finsler manifold.
The conserved current is given directly from $K^\flat$.  
We treat two examples: the Carter constant on Kerr spacetime and the 
Runge-Lentz vector in classical mechanics.

\section{Killing Symmetry on Finsler Manifold}

\noindent {\bf Definition}: A vector field $v=v^\mu  \frac{\partial}{\partial x^\mu}$ on 
Finsler manifold, of which elements $v^\mu =v^\mu (x,dx)$  are homogeneous of degree 0 with respect to $dx$, $v^\mu (x,\lambda dx)=v^\mu(x,dx)$, is called {\it generalized vector field}.  

\noindent {\bf Definition}: When the Lie derivative of Finsler metric $F$ by the 
generalized vector field $K=K^\mu (x,dx)\frac{\partial}{\partial x^\mu}$ 
satisfies the relation 
\begin{eqnarray}
{\cal L}_K F &\equiv& K^\mu \frac{\partial F}{\partial x^\mu} + dK^\mu \frac{\partial F}{\partial dx^\mu} = dB(x,dx)\ ,
\label{eq5}
\end{eqnarray}
the $K$ is called {\it quasi-Killing vector field}. When $B=0$, $K$ is just called Killing
vector field on Finsler manifold. Here $K^\mu$ and $B$ are homogeneous of degree 0 with respect to $dx$.
    
\noindent {\bf Theorem}: When $K=K^\mu(x,dx)\frac{\partial}{\partial x^\mu}$ is 
the quasi-Killing vector field satisfying ${\cal L}_K F = dB$, 
we have the conserved quantity on a solution curve ${c}$.
\begin{eqnarray}
J &\equiv& K^\mu  \frac{\partial F}{\partial dx^\mu} - B,\ \ \ \ \ {c}^*\ dJ=0\ . 
\label{eq7}
\end{eqnarray}
\noindent {\it Proof}: \ \  ${c}^*\ dJ = { c}^*\left(
dK^\mu\  \frac{\partial F}{\partial dx^\mu} 
+K^\mu\ d \frac{\partial F}{\partial dx^\mu} - dB \right) =
 {c}^*\left(
 dK^\mu \frac{\partial F}{\partial dx^\mu} 
+K^\mu \frac{\partial F}{\partial x^\mu} - dB \right) 
=$\\
$ {c}^*\left( {\cal L}_K F -dB \right) =0  $ 
on a solution curve ${c}$. Here we used the equation of motion (\ref{eq3}) in the 2nd equality.
$\qed$

Killing symmetry (or vector fields) and the corresponding conservation laws, called Noehter's theorem, on Finsler manifold are given in ref.\cite{Ootsuka}. We give their covariant (reparametrization-invariant)  form in more simple way than the usual presentation \cite{Olver}. 

\noindent {\bf Definition}: By using the non-linear connection $N^\mu{}_\alpha$, 
{\it spray operator} $\delta$
to a general function $H=H(x,dx)$ is defined by
\begin{eqnarray}
\delta H &\equiv & dx^\mu \frac{\partial H}{\partial x^\mu} - 2G^\mu \frac{\partial H}{\partial dx^\mu}\ ,\ \ \ 2G^\mu =N^\mu{}_\alpha dx^\alpha \ .
\label{eq8}
\end{eqnarray} 
The $\delta$-operation may be regarded as  a kind of variation since it is obtained by replacing $d^2x^\mu$ by $-2G^\mu$ in 
$dH=dx^\mu \frac{\partial H}{\partial x^\mu} +d^2x^\mu \frac{\partial H}{\partial dx^\mu}$.

From the definition of Finsler non-linear connection \cite{KO} given in the introduction
\begin{eqnarray} 
0 &=& \frac{\partial F}{\partial x^\mu} - N^\rho{}_\mu(x,dx) \frac{\partial F}{\partial dx^\rho},\ \ \ 
 \frac{\partial G^\rho}{\partial dx^\mu} = N^\rho{}_\mu ,\ \ 2G^\rho = N^\rho{}_\mu dx^\mu \ ,
\label{eq9}
\end{eqnarray}
we have the identity 
\begin{eqnarray}
\delta F &=& dx^\mu \frac{\partial F}{\partial x^\mu} - 2G^\rho \frac{\partial F}{\partial dx^\rho} = dx^\mu\left( \frac{\partial F}{\partial x^\mu} - N^\rho{}_\mu \frac{\partial F}{\partial dx^\rho}  \right) =0
 \ .
\label{eq10}
\end{eqnarray}  
Its derivative by $dx^\nu$ leads
\begin{eqnarray}
 0&=&\bib{F}{x^\nu}+dx^\mu \bbib{F}{dx^\nu}{x^\mu}-2\bib{G^\mu}{dx^\nu}\bib{F}{dx^\mu}
 -2G^\mu\bbib{F}{dx^\nu}{dx^\mu} \nonumber\\
  &=&-\bib{F}{x^\nu}+dx^\mu \bbib{F}{dx^\nu}{x^\mu}-2G^\mu\bbib{F}{dx^\nu}{dx^\mu}.
\label{eq11}
\end{eqnarray}
where we used Eq.~(\ref{eq9}) in the last equality.

We propose another formulation of Killing symmetry on Finsler manifold.

\noindent {\bf Definition}: Non-linear 1-form is a homogeneous function of
degree 1 with respect to $dx$. {\it Killing 1-form} is a non-linear 1-form which satisfies 
\begin{eqnarray}
\delta K^\flat  &=& 0\    .
\label{eq12}
\end{eqnarray}

\noindent {\bf Theorem}: When $K^\flat$ is Killing 1-form, $\delta K^\flat =0$, we have a conserved quantity:
\begin{eqnarray}
J &\equiv& \frac{K^\flat}{F},\ \ \ \  {c}^*\ dJ =0\ .
\label{eq13}
\end{eqnarray}
\noindent {\it Proof}:  Both $K^\flat$ and $F$ are 1st-order homogeneous function and satisfy
$\delta K^\flat =0$, $\delta F=0$.  
Thus, $J=K^\flat /F$ satisfies $\delta J=dx^\mu\frac{\partial J}{\partial x^\mu}-2G^\mu\frac{\partial J}{\partial dx^\mu}=0$, and it has homogeneity of degree 0, $dx^\mu\frac{\partial J}{\partial dx^\mu}=0$. 
Then,
${c}^*dJ={c}^*\left\{ dx^\mu\frac{\partial J}{\partial x^\mu} +  d^2x^\mu\frac{\partial J}{\partial dx^\mu} \right\}
={c}^*\left( d^2x^\mu +2G^\mu  \right)\frac{\partial J}{\partial dx^\mu}
=c^* \lambda dx^\mu \frac{\partial J}{\partial dx^\mu} =0$. $\qed$

The conserved quantity $J$ is independent of choices of parameters.
The Killing 1-form $K^\flat$  as well as the Killing vector field $K$ describe the 
conserved quantity. Both are related with each other:  

\noindent {\bf Proposition}: When Killing non-linear 1-form $K^\flat$ is represented in the form
\begin{eqnarray} 
K^\flat &=& F\left( K^\mu (x,dx) \frac{\partial F}{\partial dx^\mu} -B(x,dx) \right)
\label{eq14}
\end{eqnarray}
with some functions $K^\mu(x,dx)$ and $B(x,dx)$, 
correspondingly we obtain the quasi-Killing vector field 
as $K=K^\mu (x,dx)\frac{\partial}{\partial x^\mu}$. That is,
\begin{eqnarray} 
\delta K^\flat =0  &\Leftrightarrow& {\cal L}_K  F =dB\ \ \ {\rm on\ shell}\ ,
\label{eq15}
\end{eqnarray}
where on shell means $x^\mu$ and $dx^\mu$ are on a solution curve ${c}$, satisfying
the Euler-Lagrange equation (\ref{eq3}).

\noindent {\it Proof}:
Since $\delta F=0$,
\begin{eqnarray}
 \delta K^\flat&=&
 F\left(
 dx^\nu \bib{K^\mu}{x^\nu}\bib{F}{dx^\mu}-2G^\nu \bib{K^\mu}{dx^\nu}\bib{F}{dx^\mu}
 +K^\mu dx^\nu \bbib{F}{x^\nu}{dx^\mu}-2K^\mu G^\nu \bbib{F}{dx^\nu}{dx^\mu} 
 \right. \nonumber \\
 &&\left.
 -dx^\mu \bib{B}{x^\mu}+2G^\mu \bib{B}{dx^\mu}\right) \nonumber \\
 &=&
 F\left(
 K^\mu \bib{F}{x^\mu}+
 dx^\nu \bib{K^\mu}{x^\nu}\bib{F}{dx^\mu}-2G^\nu \bib{K^\mu}{dx^\nu}\bib{F}{dx^\mu}
  -dx^\mu \bib{B}{x^\mu}+2G^\mu \bib{B}{dx^\mu} \right) \nonumber \\
 &=&
 F\left\{
 K^\mu \bib{F}{x^\mu}+dK^\mu \bib{F}{dx^\mu}-dB
 -(d^2 x^\nu+2G^\nu) \left(
 \bib{K^\mu}{dx^\nu}\bib{F}{dx^\mu}-\bib{B}{dx^\nu}
 \right)
 \right\} \ ,
\label{eq16}
\end{eqnarray}
where by using Eq.~(\ref{eq11}) the 3rd and 4th term in RHS of the 1st equation is rewritten to the 1st term of the next.  
When on-shell $0=d^2 x^\nu+2G^\nu -\lambda dx^\nu$, and the $K^\nu$ and $B$ are homogeneity 0 functions, we obtain
\begin{eqnarray}
 \delta K^\flat=
 F\left\{ {\cal L}_K F-dB \right\}\ \ {\rm on\ shell}.
\end{eqnarray}
When $\delta K^\flat =0$ the conserved quantity $J=\frac{K^\flat}{F}$ with Eq.~(\ref{eq12}) coincides with Eq.~(\ref{eq7}) from Noether's theorem.

\section{Generalization of Killing Tensors on Finsler Manifold}

\subsection{Killing tensors}

Killing 1-form $K^\flat$ defines
the Killing tensors on Finsler manifold in very simple forms. 

\noindent {\bf Definition}: When Killing non-linear 1-form takes  
\begin{eqnarray}
 K^\flat &=& K_{\mu_1\cdots\mu_p} (x)\frac{dx^{\mu_1}\cdots dx^{\mu_p}}{F^{p-1}}\ , 
\label{eq17}
\end{eqnarray}
 the $K_{\mu_1\cdots\mu_p}(x)$ is called $p$-rank Killing tensor on Finsler manifold. 

\noindent {\bf Proposition}: 
Killing non-linear 1-form $K^\flat$ is given by $p$-rank Killing tensor,
\begin{eqnarray}
\delta K^\flat =0 & \Leftrightarrow &   \nabla_{(\mu}K_{\mu_1\cdots\mu_p)}(x)=0\ \ \ 
(,{\rm where}\ {}_{(\mu\mu_1\cdots \mu_p)}\ {\rm means\ symmetrization.})\nonumber\\
J \equiv \frac{K^\flat}{F} &=& K_{\mu_1\cdots\mu_p}(x)\frac{dx^{\mu_1}}{F}\cdots\frac{dx^{\mu_p}}{F} 
\ \ {\rm is\ conserved\ }{c}^* dJ=0\ {\rm in\ this\ case}\ ,
\label{eq17-1}
\end{eqnarray}
where $\nabla$ is covariant derivative defined by Finsler non-linear connection. 
$\nabla_\mu K_{\nu\cdots} \equiv \partial_\mu K_{\nu\cdots}-N^\rho{}_{\nu\mu}K_{\rho\cdots}-\cdots$,
where $N^\rho{}_{\mu\nu}\equiv\frac{\partial N^\rho{}_\mu}{\partial dx^\nu} = \frac{\partial^2 G^\rho}{\partial dx^\mu \partial dx^\nu}$ . 

\noindent {\it Proof}: 
\begin{eqnarray}
\delta K^\flat &=& dx^\mu \frac{\partial K^\flat}{\partial x^\mu} - 2G^\rho \frac{\partial K^\flat}{\partial dx^\rho}  \nonumber\\
 &=& \frac{dx^\mu dx^{\mu_1}\cdots dx^{\mu_p}}{F^{p-1}}\left\{  
  \frac{\partial K_{\mu_1\cdots\mu_p}(x)}{\partial x^\mu} - N^\rho{}_{\mu_1\mu}  
K_{\rho\mu_2\cdots\mu_p}(x)-\cdots -N^\rho{}_{\mu_p\mu}  
K_{\mu_1\cdots\mu_{p-1}\rho}(x)
 \right\}  \nonumber\\
 &=&  \frac{dx^\mu dx^{\mu_1}\cdots dx^{\mu_p}}{F^{p-1}} \nabla_{(\mu}K_{\mu_1\cdots\mu_p)}(x)\ ,
\label{eq18}
\end{eqnarray}
where we used $2G^\rho = N^\rho{}_{\mu\alpha}dx^\mu dx^\alpha$. 
The conservation has already been proved in Eq.~(\ref{eq13}).

\subsection{Riemannian case and Carter constant}

In Riemannian case $F=\sqrt{g_{\mu\nu}(x)dx^\mu dx^\nu}$,  
$\nabla$ becomes usual Riemannian covariant derivative and 
$N^\rho{}_{\mu\alpha}=\varGamma^\rho{}_{\mu\alpha}$.
Thus, $K_{\mu_1\cdots\mu_p}(x)$ in Eq.~(\ref{eq17}) which satisfies  $ \nabla_{(\mu}K_{\mu_1\cdots\mu_p)}(x)=0$
becomes the ordinary Killing tensors in this case. 
By taking the arc length parameter $F=d\tau$,
$J=K_{\mu_1\cdots\mu_p}(x)u^{\mu_1}\cdots u^{\mu_p}$ where $u^\mu = \frac{d x^\mu}{d\tau}=\frac{dx^\mu}{F}$.  
This is the ordinary conserved quantity made from Killing tensors along geodesics.

\noindent {\bf Proposition}:  When $K_{\mu_1\cdots\mu_p}(x)$ is 
Killing tensor on Riemannian manifold,
 $\nabla_{(\mu}K_{\mu_1\cdots\mu_p)}=0$,
\begin{eqnarray}
K &\equiv& K^\mu{}_{\nu_1\cdots\nu_{p-1}}(x)\frac{dx^{\nu_1}\cdots dx^{\nu_{p-1}}}{F^{p-1}}\frac{\partial}{\partial x^\mu}
\label{eq20}
\end{eqnarray}
 is the quasi-Killing vector field, satisfying ${\cal L}_K F =dB$ with $B=\frac{p-1}{p}J$,
where $K^\mu{}_{\nu_1\cdots\nu_{p-1}}(x)=g^{\mu\rho}(x) K_{\rho\nu_1\cdots\nu_{p-1}}(x)$.

\noindent {\it Proof}: 
\begin{eqnarray}
 {\cal L}_K F &=& {K^\mu}_{\nu_1\nu_2\dots\nu_{p-1}}
 \frac{dx^{\nu_1}dx^{\nu_2}\cdots dx^{\nu_{p-1}}}{F^{p-1}}\bib{F}{x^\mu}  \nonumber\\
&&
 + \left\{ d{K^\mu}_{\nu_1\nu_2\dots \nu_{p-1}}
 \frac{dx^{\nu_1}dx^{\nu_2}\cdots dx^{\nu_{p-1}}}{F^{p-1}} 
 +(p-1){K^\mu}_{\nu_1\nu_2\dots \nu_{p-1}}
 \frac{d^2 x^{\nu_1}dx^{\nu_2}\cdots dx^{\nu_{p-1}}}{F^{p-1}}  \right. \nonumber\\
&& \left.
\ \ \ \ \ \  -(p-1){K^\mu}_{\nu_1\nu_2\dots \nu_{p-1}}
 \frac{dx^{\nu_1}dx^{\nu_2}\cdots dx^{\nu_{p-1}}}{F^p}dF
 \right\}
 \bib{F}{dx^\mu} \nonumber \\
 &=&{K^\mu}_{\nu_1\nu_2\dots\nu_{p-1}}
 \frac{dx^{\nu_1}dx^{\nu_2}\cdots dx^{\nu_{p-1}}}{F^{p-1}}\bib{F}{x^\mu}
 + d(g^{\mu\rho}K_{\rho\nu_1\nu_2\dots \nu_{p-1}})
 \frac{g_{\mu\nu_p} dx^{\nu_1}dx^{\nu_2}\cdots dx^{\nu_{p-1}}dx^{\nu_p}}{F^{p}} 
 \nonumber \\
 && 
 +(p-1)K_{\nu_1\nu_2\dots \nu_{p}}
 \frac{d^2 x^{\nu_1}dx^{\nu_2}\cdots dx^{\nu_{p}}}{F^{p}}
 -(p-1)K_{\nu_1\nu_2\dots \nu_{p}}
 \frac{dx^{\nu_1}dx^{\nu_2}\cdots dx^{\nu_{p}}}{F^{p+1}}dF,
\label{eq21}
\end{eqnarray}
Here 
\begin{eqnarray}
 K_{\nu_1\nu_2\dots \nu_{p}}
 \frac{d^2 x^{\nu_1}dx^{\nu_2}\cdots dx^{\nu_{p}}}{F^{p}}
 &=&d\left(K_{\nu_1\nu_2\dots \nu_p}
   \frac{d x^{\nu_1} dx^{\nu_2} \cdots dx^{\nu_p}}{F^p}\right)
 -dK_{\nu_1\nu_2\dots \nu_p}\frac{d x^{\nu_1} dx^{\nu_2}\cdots dx^{\nu_p}}{F^p}
 \nonumber \\
            &-& (p-1)K_{\nu_1\nu_2\dots \nu_p}\frac{dx^{\nu_1} d^2x^{\nu_2} \cdots dx^{\nu_p}}{F^p}
 +pK_{\nu_1\nu_2\dots \nu_p}\frac{dx^{\nu_1} dx^{\nu_2}\cdots dx^{\nu_p}}{F^{p+1}}
  dF, \nonumber \\
 \therefore~
 K_{\nu_1\nu_2\dots \nu_p}\frac{d^2 x^{\nu_1} dx^{\nu_2}\cdots dx^{\nu_p}}{F^p}
 &=&d\left(\frac1p K_{\nu_1\nu_2\dots \nu_p}
   \frac{d x^{\nu_1} dx^{\nu_2}\cdots dx^{\nu_p}}{F^p}\right)
  \nonumber \\
         &-& \frac1p dK_{\nu_1\nu_2\dots \nu_p}\frac{d x^{\nu_1} dx^{\nu_2}\cdots dx^{\nu_p}}{F^p}
 +  K_{\nu_1\nu_2\dots \nu_p}\frac{dx^{\nu_1} dx^{\nu_2}\cdots dx^{\nu_p}}{F^{p+1}}dF.
\label{eq22}
\end{eqnarray}
Using the above relation,
the 3rd and 4th terms of the last equation of (\ref{eq21}) is rewritten. Then, we get
\begin{eqnarray}
 {\cal L}_K F &=&
  {K^\mu}_{\nu_1\nu_2\dots \nu_{p-1}}\frac{dx^{\nu_1}dx^{\nu_2}\cdots dx^{\nu_{p-1}}}{F^{p-1}}
  \bib{F}{x^\mu}
  + d(g^{\mu\rho}K_{\rho\nu_1\nu_2\dots \nu_{p-1}})
 \frac{g_{\mu\nu_p} dx^{\nu_1}dx^{\nu_2}\cdots dx^{\nu_{p-1}}dx^{\nu_p}}{F^{p}} 
 \nonumber \\
 &&
 -\frac{p-1}{p}dK_{\nu_1\nu_2\dots \nu_p}\frac{d x^{\nu_1} dx^{\nu_2}\cdots dx^{\nu_p}}{F^p}
 +d\left(\frac{p-1}{p} K_{\nu_1\nu_2\dots \nu_p}
   \frac{d x^{\nu_1} dx^{\nu_2}\cdots dx^{\nu_p}}{F^p}\right)
  \nonumber \\
 &=&
 {K^\mu}_{\nu_1\nu_2\dots \nu_{p-1}} \frac12 \bib{g_{\nu_0\nu_p}}{x^\mu}
 \frac{dx^{\nu_0} dx^{\nu_1}dx^{\nu_2}\cdots dx^{\nu_{p}}}{F^{p}}
 -g^{\mu\rho}dg_{\mu\nu_p}K_{\rho\nu_1\nu_2\dots \nu_{p-1}}
 \frac{dx^{\nu_1}dx^{\nu_2}\cdots dx^{\nu_{p-1}}dx^{\nu_p}}{F^{p}} 
 \nonumber 
 \\
 &&
 +\frac1p dK_{\nu_1\nu_2\dots \nu_p}\frac{d x^{\nu_1} dx^{\nu_2}\cdots dx^{\nu_p}}{F^p}
 +d\left(\frac{p-1}{p}J\right)
 \nonumber 
 \\
 &=&
 \left\{
 {K^\mu}_{\nu_1\nu_2\dots \nu_{p-1}} \varGamma_{\nu_0\nu_p\mu}
 -g^{\mu\rho}(\varGamma_{\mu\nu_p\nu_0}+\varGamma_{\nu_p\mu\nu_0})
 K_{\rho\nu_1\nu_2\dots \nu_{p-1}}
 +\frac1p \bib{K_{\nu_1\nu_2\dots \nu_p}}{x^{\nu_0}}
 \right\}
 \frac{dx^{\nu_0}dx^{\nu_1}\cdots dx^{\nu_p}}{F^{p}} 
 \nonumber 
 \\
 && +d\left(\frac{p-1}{p}J\right)
 \nonumber 
 \\
 &=&
 \frac1p \left\{
 \bib{K_{\nu_1\nu_2\dots \nu_p}}{x^{\nu_0}}
 -p K_{\rho\nu_1\nu_2\dots \nu_{p-1}}{\varGamma^\rho}_{\nu_p\nu_0}
 \right\} 
 \frac{dx^{\nu_0}dx^{\nu_1}\cdots dx^{\nu_p}}{F^{p}}
 +d\left(\frac{p-1}{p}J\right) \nonumber\\
 &=&  \frac{1}{p} \nabla_{(\nu_0} K_{\nu_1\cdots\nu_p)}  
 \frac{dx^{\nu_0}dx^{\nu_1}\cdots dx^{\nu_p}}{F^{p}}
 +
d\left(\frac{p-1}{p}J\right)
\label{eq23}
\end{eqnarray}
Since $K_{\nu_1\cdots\nu_p}$ satisfies $\nabla_{(\nu_0} K_{\nu_1\cdots\nu_p)}=0$,
the generalized vector Eq.~(\ref{eq20}) is the quasi-Killing vector, satisfying ${\cal L}_K F=dB$, even without using on-shell condition.

Carter constant is the conserved quantity along the geodesic motion of point particle under Kerr spacetime. 
It is derived from the 2nd-rank Killing tensor 
in Riemannian geometry.
The Kerr metric is simply given by using vierbein as
\begin{eqnarray}
g_{\mu\nu}(x)dx^\mu dx^\nu &=& -(\theta^1)^2-(\theta^2)^2-(\theta^3)^2 +(\theta^0)^2
\label{eq24}\\
\theta^1 = \frac{\rho}{\sqrt\Delta}dr, &\ \ & \theta^2=\rho d\theta,\ \ \theta^3=\frac{S}{\rho}\left\{  
(r^2+a^2)d\phi - a dt\right\},\ \ \theta^0=\frac{\sqrt\Delta}{\rho}\left( dt - aS^2 d\phi \right)
\ \ \ \ 
\label{eq25}
\end{eqnarray}
where $\rho^2=r^2 +a^2C^2$, $\Delta=r^2+a^2-2Mr$, and $C,S={\rm cos}\theta,{\rm sin}\theta$.
The Killing tensor, denoted as $K_{\mu\nu}^{\rm Carter}$, 
is read off from the Carter constant \cite{Carter,ONeill} as
\begin{eqnarray}
K^\flat &=& K_{\mu\nu}^{\rm Carter}(x)\frac{dx^\mu dx^\nu}{F}  =
   a^2C^2 \frac{1}{F}\left\{ (\theta^0)^2-(\theta^1)^2 \right\} + r^2 \frac{1}{F}\left\{ (\theta^2)^2+(\theta^3)^2 \right\}
\label{eq26}
\end{eqnarray} 
The quasi-Killing vector field corresponding to the Carter constant is given by 
\begin{eqnarray}
 K^{\rm Carter} &=& g^{\mu\rho}K_{\rho\nu}^{\rm Carter} \frac{dx^\nu}{F}\bib{}{x^\mu}\ ,
\label{eq27}
\end{eqnarray}
which satisfies, from the last proposition, the relation (\ref{eq23}) with $p=2$ as 
\begin{eqnarray}
 {\cal L}_{K^{\rm Carter}} F &=& d\left(\frac{1}{2}  K^{\rm Carter}_{\mu\nu}\frac{dx^\mu}{F}\frac{dx^\nu}{F}
 \right)\ .
 \label{eq28}
\end{eqnarray}
We checked the above equation is actually derived from Lie derivative of the vector field Eq.~(\ref{eq27}) without using on-shell condition. This shows $\nabla_{(\lambda}K^{\rm Carter}_{\mu\nu )}=0 $.
The generalized Killling vector (\ref{eq27}) represents a hidden symmetry \cite{Olver} on Kerr 
spacetime.

\subsection{Runge-Lentz vectors in classical mechanics}

We can invent a way of finding the Runge-Lentz vectors in Newtonian system,
by using the Killing non-linear 1 form $K^\flat$. 
Classical Lagrangian system is treated in Finsler geometry
with the metric 
\begin{eqnarray}
F(x,dx) &=& \frac{m}{2}\frac{(dx^1)^2+(dx^2)^2+(dx^3)^2}{dx^0} + \frac{GMm}{r}dx^0
\label{eq29}
\end{eqnarray}
which satisfies the 1st-order homogeneity relation $F(x,\lambda dx)=\lambda F(x,dx)$.
The equation (\ref{eq4-2}) leads the 
non-linear connection~\cite{KO}
\begin{eqnarray}
2G^0 = -\frac{2GMm}{r^2}\frac{(dx^0)^2dr}{F},&\ \ &\ \ \ \ \  
2G^a = - \frac{2GMm}{r^2}\frac{dx^0 dx^a dr}{F} + \frac{GMx^a}{r^3}(dx^0)^2
\label{eq30}
\end{eqnarray}
with space-component $a=1,2,3$. 
We adopt a different ansatz of the 2nd-rank Killing tensor $K_{\mu\nu}(x)$ from Eq.~(\ref{eq17}),
\begin{eqnarray}
K^\flat &=& F K_{\mu\nu}(x)\frac{d x^\mu}{dx^0}\frac{dx^\nu}{dx^0}\ ,
\label{eq31}
\end{eqnarray}
 by considering the non-relativistic nature of the system.
Then,
\begin{eqnarray}
 \delta K^\flat &=&
F\left[ 
   dx^\lambda \frac{\partial K_{\mu\nu}(x)}{\partial x^\lambda} \frac{dx^\mu}{dx^0} \frac{dx^\nu}{dx^0}
 -2G^\lambda \frac{\partial}{\partial dx^\lambda} \left\{ \frac{dx^\mu}{dx^0}\frac{dx^\nu}{dx^0}   \right\} K_{\mu\nu}(x)
  \right]  \nonumber\\
&=& F \left[ \frac{\partial K_{\mu\nu}}{\partial x^\lambda}\frac{dx^\lambda dx^\mu dx^\nu}{(dx^0)^2}
      -\frac{GMx^a}{r^3} (dx^0)^2 \frac{\partial}{\partial dx^a} \left\{ \frac{dx^\mu}{dx^0}\frac{dx^\nu}{dx^0}   \right\} K_{\mu\nu}(x)
    \right]  \nonumber\\
&=&  F \left[ \frac{\partial K_{\mu\nu}}{\partial x^\lambda} \frac{dx^\lambda dx^\mu dx^\nu}{(dx^0)^2}
      - \frac{2GMx^a}{r^3} K_{a\nu}dx^\nu
    \right]  \ , \ \ \ \ \ \ 
 \label{eq31}
\end{eqnarray}
where in the 2nd equality we used the fact that $2G^0$ and the 1st term of $2G^a$ in Eq.~(\ref{eq30}) cancel since
$J$ in $K^\flat = FJ$ is homogeneous of degree 0.
The symmetry condition $\delta K^\flat=0$ requires the relations
\begin{align}
&\frac{\partial K_{00}}{\partial x^0}-2K_{0a}\frac{GMx^a}{r^3} =0 
   & \frac{\partial K_{ab}}{\partial x^0} + \frac{\partial K_{0a}}{\partial x^b}+ \frac{\partial K_{0b}}{\partial x^a}&=0    \nonumber\\
&\frac{\partial K_{00}}{\partial x^a}+2\frac{\partial K_{0a}}{\partial x^0}-2K_{ab}\frac{GMx^b}{r^3}=0  
 & \frac{\partial K_{ab}}{\partial x^c}+ \frac{\partial K_{bc}}{\partial x^a}+ \frac{\partial K_{ca}}{\partial x^b}  & =3\partial_{(c}K_{ab)}=0\ .
\label{eq32}
\end{align}
By solving these equations we can find a hidden conserved quantity.
Requiring the static condition, $\frac{\partial K_{\mu\nu}}{\partial x^0}=0$, 
$K_{0a}=0$ from the 1st equations. The 2nd equations are automatically satisifed.
The point is the 4th equations $\partial_{(c}K_{ab)}=0$. There are
$_3H_3=10$ equations.
By taking the tensorial form $K_{ab}=\delta_{ab}f({\bm x}) +x^a x^b g({\bm x})$, we have obtained the 
energy and total angular momentum. Thus, we must seek for non-tensorial form.
$K_{11}=K_{11}(x^2,x^3)$ since $\partial_1 K_{11}=0$.
$\partial_2 K_{11}+2\partial_1 K_{12}=0$ and $\partial_1 K_{22}+2\partial_2 K_{12}=0$ leads
$(K_{11},K_{12},K_{22})=(-f(x^2),\frac{x^1}{2}f^\prime (x^2),\frac{(x^1)^2}{2}f^{\prime\prime}(x^2))$
with arbitrary function $f(x^2)$. The  simplest, non-trivial solution is 
 $(K_{11},K_{12},K_{22})=c_1(-x^2,\frac{x^1}{2},0)$ with constant $c_1$. There are 6 independent solutions of this type.
Then, the solutions with 6 constants $c_{1-6}$ are 
 $(K_{11},K_{22},K_{33};K_{12},K_{13},K_{23})
=(-c_1x^2-c_2x^3,-c_3x^3-c_4x^1,-c_5x^1-c_6x^2;
c_1\frac{x^1}{2}+c_4\frac{x^2}{2}, c_2\frac{x^1}{2}+c_5\frac{x^3}{2},c_3\frac{x^2}{2}+c_6\frac{x^3}{2})$.
The 3rd equation $\frac{\partial K_{00}}{\partial x^a} = 2K_{ab}x^b \frac{GM}{r^3}$ is regarded as 
an integrability condition requiring outer derivative of $dx^a \partial_a K_{00}$ should vanish.
It leads three solutions:
\begin{align}
(c_4=c_5=1)  && K_{00}=\frac{GMx^1}{r}  && K_{22}=K_{33}=-x^1    && K_{12}=\frac{x^2}{2}   && K_{13}=\frac{x^3}{2}\nonumber\\
(c_6=c_1=1)  && K_{00}=\frac{GMx^2}{r}  && K_{33}=K_{11}=-x^2    && K_{23}=\frac{x^3}{2}   && K_{12}=\frac{x^1}{2}\nonumber\\
(c_2=c_3=1)  && K_{00}=\frac{GMx^3}{r}  && K_{11}=K_{22}=-x^3    && K_{13}=\frac{x^1}{2}   && K_{23}=\frac{x^2}{2}
\label{eq33}
\end{align}
where the other $c_i$'s and $K_{\mu\nu}$'s are 0 in each line.
These are called Runge-Lentz tensors which lead Runge-Lentz vectors through the standard procedure
$J=\frac{K^\flat}{F}=K_{\mu\nu} \frac{dx^\mu}{dx^0}\frac{dx^\nu}{dx^0}$.
They are denoted as $J_{1,2,3}$ in order, which are given by
\begin{eqnarray}
J_1 &=& \frac{GM m x^1}{r} - mx^1\left\{ \left(\frac{dx^2}{dx^0}\right)^2 + \left(\frac{dx^3}{dx^0}\right)^2  \right\} + m \frac{dx^1}{dx^0}\left( x^2\frac{dx^2}{dx^0} + x^3\frac{dx^3}{dx^0}  \right) 
\nonumber\\
J_2 &=& \frac{GM m x^2}{r} - mx^1\left\{ \left(\frac{dx^3}{dx^0}\right)^2 + \left(\frac{dx^1}{dx^0}\right)^2  \right\} + m \frac{dx^2}{dx^0}\left( x^3\frac{dx^3}{dx^0} + x^1\frac{dx^1}{dx^0}  \right) 
\nonumber\\
J_3 &=& \frac{GM m x^3}{r} - mx^1\left\{ \left(\frac{dx^1}{dx^0}\right)^2 + \left(\frac{dx^2}{dx^0}\right)^2  \right\} + m \frac{dx^3}{dx^0}\left( x^1\frac{dx^1}{dx^0} + x^2\frac{dx^2}{dx^0}  \right)
\label{eq34}
\end{eqnarray}
where we took $m$ as the overall proportional constant.
Runge-Lentz vectors are also derived, similarly to our method, in ref.\cite{Stephani} 
using Hamiltonian formalism. 
Finally we give the generalized Killing vectors corresponding to Eq.~(\ref{eq33}). We take
\begin{eqnarray}
K &=& K^a(x,dx) \frac{\partial}{\partial x^a} = 2K_{ab}(x)\frac{dx^b}{dx^0} \frac{\partial}{\partial x^a}\ .
\label{eq35}
\end{eqnarray}
which gives three quasi-Killing vector fields corresponding to three lines of (\ref{eq33}):
\begin{eqnarray}
K_1 &=& \left( x^2\frac{dx^2}{dx^0} + x^3\frac{dx^3}{dx^0} \right)\frac{\partial}{\partial x^1}
 +  \left( x^2\frac{dx^1}{dx^0} - 2x^1\frac{dx^2}{dx^0} \right)\frac{\partial}{\partial x^2}
 +  \left( x^3\frac{dx^1}{dx^0} - 2x^1\frac{dx^3}{dx^0} \right)\frac{\partial}{\partial x^3}\nonumber\\
K_2 &=&  \left( x^1\frac{dx^2}{dx^0} - 2x^2\frac{dx^1}{dx^0} \right)\frac{\partial}{\partial x^1}
 + \left( x^3\frac{dx^3}{dx^0} + x^1\frac{dx^1}{dx^0} \right)\frac{\partial}{\partial x^2}
 + \left( x^3\frac{dx^2}{dx^0} - 2x^2\frac{dx^3}{dx^0} \right)\frac{\partial}{\partial x^3} \nonumber\\
K_3 &=& \left( x^1\frac{dx^3}{dx^0} - 2x^3\frac{dx^1}{dx^0} \right)\frac{\partial}{\partial x^1}
 + \left( x^2\frac{dx^3}{dx^0} - 2x^3\frac{dx^2}{dx^0} \right)\frac{\partial}{\partial x^2}
 + \left( x^1\frac{dx^1}{dx^0} + x^2\frac{dx^2}{dx^0} \right)\frac{\partial}{\partial x^3}.\ \ \ \ \ \ \ \ \ \ 
\label{eq36}
\end{eqnarray}
They satisfy the relation
\begin{eqnarray}
{\cal L}_K F &=& dB,\ \ \ B = -mK_{00} + K_{ab}\frac{dx^a}{dx^0}\frac{dx^b}{dx^0}\  ,
\label{eq37}
\end{eqnarray}
and represent hidden symmetries of the system.
The corrresponding conserved currents
$J=K^a\frac{\partial F}{\partial dx^a} -B = m \left( K_{00}+K_{ab}\frac{dx^a}{dx^0}\frac{dx^b}{dx^0}  \right)$ 
lead Eq.~(\ref{eq34}).

\section{Concluding remarks}

We have formulated the Killing vector field $K$ and the conserved quantity 
on Finsler manifold in a covariant form.
The Killing symmetry on the Finsler manifold is represented in another form by using the 
Killing 1-form $K^\flat$, which is invented for the purpose of giving Killing tensors simply. 
We have shown that Killing tensors are represented as symmetries on Finsler manifold,
as well as Killing vectors.
Using the spray operator $\delta$ and the non-linear connection~\cite{KO}, the condition $\delta K^\flat =0$ is proved to be equivalent to
the conservation law ${c}^*\ d \left(\frac{K^\flat}{F}\right)=0$ on the solution curve
 ${c}$. They are generalization of higher derivative conserved quantities
on Finsler manifold, which may be called ``hidden" conserved quantities.
 The Runge-Lentz vectors are of this kind.
We have invented an analytical way of finding these quantites.
The equation $\delta K^\flat =0$ leads partial differential equations. By solving them
we obtain the hidden conserved quantities straightforwardly 
through our ansatz of the Killing tensors.
This method gives a new analytical way of finding 
the conserved quantity which cannot be found easily in the standard Noether procedure.
 
\begin{acknowledgements}
We thank Prof. G.~W.~Gibbons, Prof. E.~Tanaka, Prof. T.~Aikou, and Prof. M.~Morikawa for valuable discussions.
T. Ootsuka thanks JSPS Institutional Program for Young Researcher Overseas Visits. 
\end{acknowledgements}

\end{document}